\documentclass[traditabstract]{aa}  
\usepackage{xcolor}
\usepackage{txfonts}
\usepackage{graphicx}
\usepackage{epsfig}
\usepackage{natbib}
\usepackage{url}
\usepackage{twoopt}
\usepackage{dcolumn}
\usepackage[colorlinks=true]{hyperref}
\hypersetup{citecolor=blue}

\begin{document} 

\title{New theoretical study of potassium perturbed by He
 and a comparison to laboratory spectra
      \thanks{K-He opacity tables for the D1 and D2 components of the
       resonance line are only available at the CDS
    via anonymous ftp to cdsarc.u-strasbg.fr (130.79.128.5) }}

   \subtitle{}

   \author{N. F. Allard         \inst{1,2}
     \and J. F. Kielkopf        \inst{3}  
     \and K. Myneni             \inst{4}
     \and J. N. Blakely         \inst{4}
}     

   \institute{GEPI, Observatoire de Paris, PSL Research University, 
     UMR 8111, CNRS, Sorbonne Paris Cit\'e, 
     61, Avenue de l’Observatoire, F-75014 Paris, France\\
              \email{nicole.allard@obspm.fr}
         \and
         Institut d'Astrophysique de Paris,  UMR7095, CNRS, 
         Universit\'e Paris VI, 98bis Boulevard Arago, F-75014 PARIS, France \\
          \and
          Department of Physics and Astronomy, 
          University of Louisville, Louisville, Kentucky 40292 USA \\       
         \and
         U.S. Army DEVCOM, Aviation and Missile Center, Redstone Arsenal,
         AL 35898 USA \\
}

   \date{ / Received: 22 November 2023/ Accepted: 20 December 2023}

\abstract{
  The visible and near-infrared spectra of late  L- and T-type
  dwarf stars are dominated in large part by the resonance lines of neutral
  Na and K. It is the collision broadening of these atomic lines by
 H$_2$ and He in the stellar atmosphere that
  determines the continuum from below 0.5$~\mu$m to above
  0.9~$\mu$m in the spectrum.
Their line profiles can be detected as far as 3000~cm$^{-1}$ from the line
center in T dwarfs and consequently an accurate and detailed determination of the complete
profile, including the extreme far wing, is required to model
the contribution of these strong alkali resonance lines to brown dwarf spectra. 
We report on our new calculations of unified
line profiles of K perturbed by He using ab initio potential data
for the conditions prevailing in cool substellar brown dwarfs and hot dense planetary atmospheres with temperatures
from $T_\mathrm{eff}$=500~$\ \mathrm{K}$ to 3000~$\ \mathrm{K}$.
For such objects with atmospheres of H$_2$ and He,  
conventional laboratory absorption spectroscopy can be used to examine the line wings
and test the line shape theories and molecular potentials. 
We find that an analytical Lorentzian profile is useful for a few cm$^{-1}$ from the line center, but not in the line wings, where the radiative transfer is a consequence of the K-He radiative collisions that are sensitive to the interaction potentials.
 Tables of the K--He absorption coefficients of the resonance lines allow accurate model atmospheres and synthetic spectra. For this purpose, we present new opacities from comprehensive line shape theory incorporating accurate ab initio potentials. Use of these new tables for the modeling of emergent spectra will be an improvement over previous line shape approximations based on incomplete or inaccurate potentials. We also present Lorentzian impact parameters obtained in the semi-classical and quantum
theory for the K $4s-4p$ resonance line centered
at 0.77~$\mu$m specifically for the line core regime.
}


   \authorrunning{N.~F.~Allard et al}

   \titlerunning{New theoretical study of potassium perturbed by He
 and comparison to laboratory spectra}
   
   \maketitle
%

   \section{Introduction}
   
   In earlier phases of the evolution of a brown dwarf star, 
most refractory metals have condensed to grains that have settled below their now
fully radiative photosphere. The alkali 
elements bind less easily to molecules or grains, and their resonance transitions remain the last sources of optical opacity.
 The  importance  of  the  far  wings  of  the
potassium doublet, centered on 0.77~$\mu$m in the spectra of methane brown
dwarfs, has been demonstrated by \citet{burrows2000}.
The interatomic interactions of the low-lying states of the alkali atom
with molecular hydrogen and helium are the main physical
quantities needed for a good understanding of collisional processes.
Calculations of \citet{allard2001} included Lorentzian
profiles using the van der Waals damping constant 
to generate atmosphere
models of brown dwarfs that underestimated the observed line strength out
to 0.9~$\mu$m for the K lines but produced too strong absorption farther from the core.
This motivated \citet{burrows2002} to modify
Lorentzian profiles and to introduce  cutoffs and other parameters that lack theoretical foundation.
Now that theoretical potentials for the binary interactions of alkalies
perturbed by He and H$_2$ may be computed with high accuracy,  
the solution to the radiative collision problem may be founded on an
appropriate theoretical framework for the line shape from first principles.

A subsequent improvement   by \citet{burrows2003}
 used multiconfiguration self-consistent
  field Hartree-Fock potentials in the \citet{szudy1975,szudy1996}
  line shape approximation, and in \citet{allard2003}  we presented absorption profiles 
of sodium and potassium perturbed by He and H$_2$ calculated
in a semi-classical (SC) unified line shape theory \citep{allard1999} using pseudo-potentials
of \citet{pascale1983} and \citet{rossi1985} for, respectively,  alkali--He
and alkali--H$_2$ interactions.
These  line profiles  were included in model atmospheres and synthetic
spectra using the \citep{allard2001} stellar atmosphere program PHOENIX.
The results were compared to previous models in Fig.~4 of \citet{allard2003},
where the observed spectrum of the methane brown dwarf Gliese 229 was plotted
for comparison. The new profiles exhibited significantly more opacity within
the first 1200~\AA\/ from the line center and less opacity
further out in the red wings, as was found in the observed spectra. The optical pseudo-continuum was therefore
depressed, while raised at the flux maximum near 1.1~$\mu$m compared to a
model based on the van der Waals approximation. The detectability of
a K--H$_2$ line satellite was predicted as well. 

In a continuation of  \citet{allard2023},  one  goal of this paper is  to
update K--He opacity tables and line parameters of the resonance line
\citep{allard2007c}  previously based on the use of the pseudo-potentials of \citet{pascale1983}.
For this purpose, the Na  and  K  line profiles perturbed by helium need to be calculated
using  up-to-date molecular data that affect the blue line wing  
satellite frequency; that is, the Na--He line satellite on the short wavelength side of the line core is closer to the unperturbed line center
than was obtained with \citet{pascale1983} (see Fig.~4 of \citet{allard2023}).  
Previous opacity tables  were constructed that enabled the computation of line profiles
for the $D1$ and $D2$ lines of Na and K  broadened by collisions with He
up to $n_{\mathrm{He}}$=$10^{19}$~cm$^{-3}$.
New opacity tables of Na--He are now  archived at the CDS and are the basis of 
 line profiles for the $D1$ and $D2$ components
 to $n_{\rm He}$=$10^{21}$~cm$^{-3}$ 
 from $T_\mathrm{eff}$=150$\ \mathrm{K}$ to 2500~$\ \mathrm{K}$ and to
 $n_{\rm He}$=$10^{22}$~cm$^{-3}$ 
 from $T_\mathrm{eff}$=3000$\ \mathrm{K}$ to 10000~$\ \mathrm{K}$.
 Accurate pressure-broadened profiles that are valid at very
 high densities of He are required for use in spectral models
 of cool white dwarf stars \citep{blouin2019c}.
 New K--H$_2$  \citep{allard2016b} and Na--H$_2$ \citep{allard2019} tables
 that reach higher densities than  $n_{\rm H2}$=$10^{19}$ cm$^{-3}$ 
are valuable and  are now used in many studies of brown dwarfs
  \citep[e.g.,][]{marley2017,oreshenko2020,lacy2023}
  and exoplanets
  \citep[e.g.,][]{phillips2020,changeat2020,freedman2021,chubb2021,nikolov2022,chubb2023}.
  Laboratory experiments serve to test and refine the theoretical models,
  the weak point of which is the knowledge of the interaction potential for
  the perturbing He atom or molecular hydrogen with the alkalies.
  Here we adopt a combination of the ab initio  potentials of 
\citet{santra2005} and \citet{nakayama2001b} (Sect.~\ref{sec:pot})   
and the dipole moments of  \citet{santra2005} for the resonance line. 
In  Sect.~\ref{sec:satellites} 
we illustrate the evolution of the absorption
spectra of K--He collisional profiles for the densities
and temperatures prevailing in the atmospheres of brown dwarf stars.
These calculations span the range
$T_\mathrm{eff}$=500$\ \mathrm{K}$ to 3000~$\ \mathrm{K}$ and
take into account  the spin-orbit coupling as described
by  \citet{allard2006}.
Spectroscopic measurement of the far line wing is a sensitive tool for
examining potentials  (Sect.~\ref{sec:exp}).
 The K--H$_2$ laboratory results reported in  \citet{allard2016b}  confirm that
the identification of a brown dwarf spectral feature previously interpreted as
CaH absorption is the line satellite of K perturbed by H$_2$. 
An additional contribution from the K--He opacity may improve the
agreement with the observation  (Sect.~\ref{sec:Tdwarf}).
We compare the obtained spectral line parameters  to those calculated
   using the quantum Baranger-Lindholm theory (Sect.~\ref{sec:param}).   
Impact broadening and shift, and satellites in the line wings due to binary
collisions, are all sensitive to different details of the atomic and
molecular interactions, as is shown in the appendices.

\section{K--He unified profiles}

In our work, a unified line shape theory and a set of atomic interaction
potential energies were used to model the entire line profile from the
impact-broadened line center to the far wing.
Complete details and the derivation of the theory 
are given in~\citet{allard1999}. A rapid account of the theory is given
in a recent paper \citep{allard2023}.

Prominent line satellites with He and H$_2$ 
are excellent tests of line shape theory and atomic interaction
potentials \citep{allard2012b,allard2016b,allard2019,allard2023}. 

\subsection{K--He diatomic potentials}
\label{sec:pot}

The theoretical  potentials  for the  binary
interactions of  alkali atoms perturbed  by He were computed
for  the  lower  states  to   an  accuracy  suitable  for  line  shape
calculations  by \citet{pascale1983}.
They were obtained by using $l$-dependent pseudo-potentials
with parameters constrained by spectroscopic and scattering data.
These potentials were used for the resonance lines of
Na, K~\citep{allard2003},
Li~\citep{allard2005}, Rb, and Cs~\citep{allard2006}.

\begin{figure}
\resizebox{0.46\textwidth}{!}{\includegraphics*{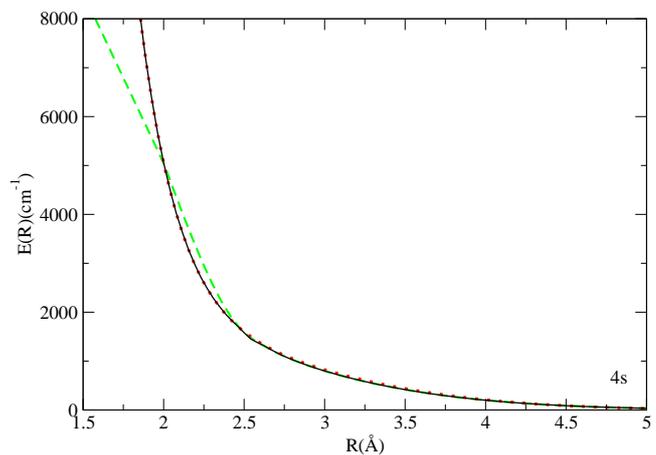}}
\caption  {Potential curve for the  4$s$ $X$ $^2\Sigma_{1/2}$ 
   state of K-He used in this work (black line).
The K-He potentials  of~\citet{nakayama2001b} (dashed green line) and  
~\citet{santra2005} (dotted red line) are superimposed. 
 \label{potx}}
\end{figure}

\begin{figure}
\resizebox{0.46\textwidth}{!}{\includegraphics*{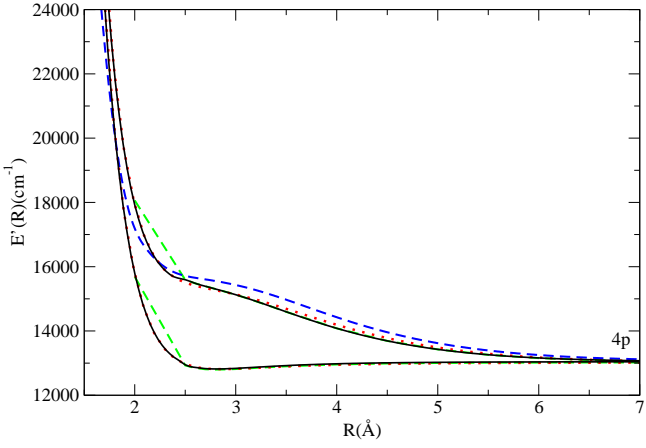}}
\caption  {Potential curves for the
  4$p$ $A$ $\Pi$ and 4$p$ $B$ $^2\Sigma$ states of the
  K-He molecule for ab initio potentials of
  \citet{nakayama2001b} (dashed green line) are compared to potentials of 
  \citet{santra2005}  (dotted red line), 
  pseudo-potentials of \citet{pascale1983} for the
  $B$ $^2\Sigma$ state (dashed blue line), and the adopted potentials (black line).
\label{potKHe}}
\end{figure}

However,  with vast improvements in computational power and refinements
in code and methodology,  significant progress in the ab initio
potentials  of the resonance line of K
has been been achieved, as was reported by  \citet{nakayama2001b},
\citet{enomoto2004}, \citet{santra2005}, \citet{mullamphy2007},
\citet{alioua2012}, and \citet{blank2012}.
In the present work, we used the K--He ab initio  potentials of 
\citet{santra2005} at a short internuclear distance and those of
\citet{nakayama2001b} elsewhere. 
Figures~\ref{potx}-~\ref{potKHe} show our adopted potentials (black line)
compared to the potential data of \citet{nakayama2001b} (dashed green line)
and to those of \citet{santra2005} (dotted red line).
In Fig.~\ref{potKHe} we also compare our potentials with those computed by~\citet{pascale1983} (dashed blue line).
The main difference occurs at short distances in the 
repulsive wall of the $B$  potential of ~\citet{pascale1983}
that is less repulsive. This difference affects the blue satellite position,
as is described in the next section.

Laboratory experiments reported in  \citet{kielkopf2012}
and \citet{kielkopf2017} served to test  the theoretical line shape
calculations based on ab initio potentials of \citet{santra2005}
to determine the  blue wing of the  potassium resonance lines
broadened by He. 
The improvement over our previous work consists of more accurate
 intermediate and long-range parts of the K--He potential
curves achieved by  \citet{nakayama2001b}, which enables a
 better determination of the line parameters of the two components of the 
doublet (see Sect.~\ref{sec:param}).

\begin{figure}
 \centering
\resizebox{0.46\textwidth}{!}
{\includegraphics*{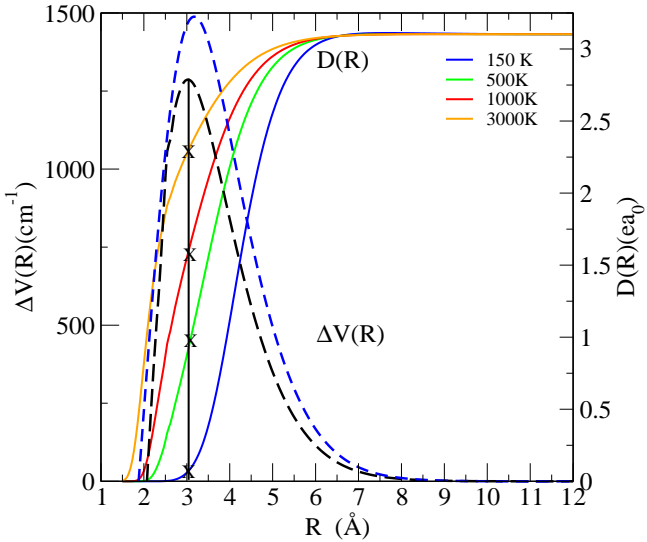}}
\caption  {$\Delta V(R)$  (dashed black line) compared to 
\citet{pascale1983} (dashed blue line) and the temperature dependence of 
the modulated dipole, $D(R)$, (solid line)
 corresponding to the 
  \mbox {4$s$ $X$    $\rightarrow$ 4$p$ $B$   $P_{3/2}$} transition
of the K $D2$ line. }
\label{diffpotKHe}
\end{figure}

\subsection{Quasi-molecular absorption in the blue wing}
\label{sec:satellites}

The line wing generally does not decrease  monotonically with increasing
frequency separation from the line center. Its shape, for an atom in the
presence of other atoms,
is sensitive to the difference between the initial   and
final state interaction potentials.
When this difference, for a given transition, goes through an
extremum, a wider range of interatomic distances contribute to the
same spectral frequency, resulting in an enhancement, or satellite, in
the line wing. Satellites in alkali spectra have been known 
since the 1930s \citep{allard1982}. More recently, laboratory spectra of
both Na and K alkalies with H$_2$, He, and other 
rare gases have been found to exhibit a systematic pattern of satellites 
in the blue wings. Quasi-molecular satellites are associated with each gas
and their presence is predicted by line shape theory using the most accurate
atomic potentials available (see Fig.3 of \citet{kielkopf2017}).
The line wing intensities are most sensitive to the 
values of the difference potential at relatively short internuclear distances, which is why they require the use of accurate atomic potentials.
Blue satellite bands in alkali-He/H$_2$ profiles
are correlated with maxima in the excited $B$
state potentials and can be predicted from the  maxima in 
the difference potentials, $\Delta V$, for the $B$-$X$ transition.

In the case of K--He, the 
$4s\;^2\Sigma - 4p\;^2\Sigma$ transition goes through a maximum, $\Delta V$, when
the atoms are 3~\AA\/ apart (Fig.~\ref{diffpotKHe}).
This leads to a ``blue'' satellite on the short
wavelength wing of the resonance doublet.
The difference potential maxima, as is shown in
 Fig.~\ref{diffpotKHe}, are 1480 and 1290~cm$^{-1}$
for the $B$--$X$  transition when using, respectively, the pseudo-potentials of 
\citet{pascale1983} and our adopted ab initio potentials.
Far wings are extended to more than 2000~cm$^{-1}$ from the
  line center when the helium density is 10$^{20}$~cm$^{-3}$
  at 3000$\ \mathrm{K}$. Fig.~\ref{extension} shows  how the unified
  theoretical approach is a major improvement compared to the unrealistic use
  of a  Lorentzian so far in  the wings.

\begin{figure}
 \centering
\resizebox{0.46\textwidth}{!}
{\includegraphics*{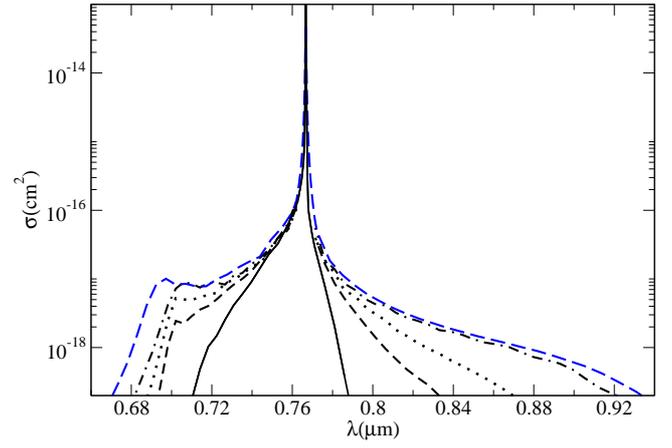}}
\caption  {Variation in the absorption cross section 
of the $D2$  component with temperature,
(from top to bottom, \mbox{$T$  =3000, 1000, 500, and 150$\ \mathrm{K}$}, 
n$_{\rm He}$=10$^{20}$~cm$^{-3}$). The corresponding  profile for
$T$=3000$\ \mathrm{K}$ using the pseudo-potentials of \citet{pascale1983}
is overplotted (dashed blue line).}
\label{varTD2}
\end{figure}

\begin{figure}
  \centering
 \resizebox{0.46\textwidth}{!}
{\includegraphics*{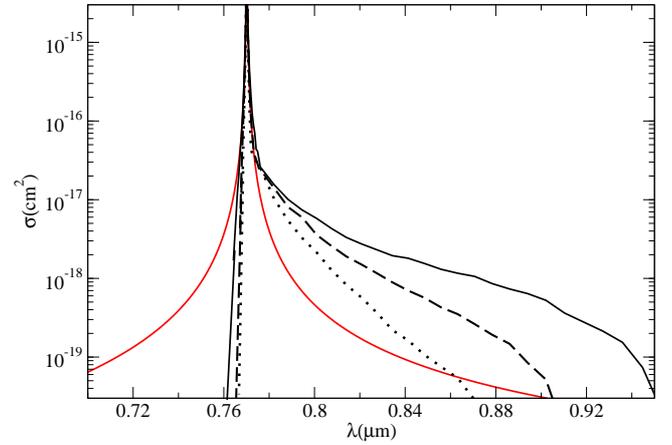}} 
\caption  {Variation in the absorption cross section of the $D1$
  component with temperature (from top to bottom, \mbox{$T$  =3000, 1000, and  500$\ \mathrm{K}$},
  n$_{\rm He}$=10$^{20}$ cm$^{-3}$). }
\label{varTD1}
 \end{figure}

Another important factor for the presence of spectral line satellites
is the variation in the electric dipole transition moment during
the collision, modulated by the Boltzmann factor, $e^{-\beta V_e(R)}$.
Here, $V_e$ is the ground state potential when we consider absorption profiles, 
or an excited state for the calculation of a profile in emission.
In Eq.~117 of \citet{allard1999}, we define 
$\tilde{d}_{ee'}(R(t))$ as a modulated dipole,
\begin{equation}
D(R) \equiv \tilde{d}_{ee'}[R(t)] = 
d_{ee'}[R(t)]e^{-\frac{V_{e}[R(t)]}{2kT} } \; , \;
\label{eq:dip} 
\end{equation}

where $d(R)$ is the transition dipole moment of  \citet{santra2005}.
 The presence of line satellite features is very sensitive to the  temperature,
due to the fast variation of the  modulated dipole moment,  $D(R$)
(Eq.~(\ref{eq:dip})), with temperature (Fig.~\ref{diffpotKHe}).  The magnitude
of the dipole moment determines the relative significance these regions
have within the spatial volume where collisions contribute to the line wing.
However, the modulated dipole decreases significantly inside 4~\AA\/, and
the region in which the satellite would develop contributes only at elevated
temperatures larger than 500$\ \mathrm{K}$.
The behavior for Na with He is similar \citep{allard2023}. 
In Fig.~\ref{varTD2} we illustrate the evolution of  the
absorption cross section of the resonance line of K for a
He density of  10$^{20}$ cm$^{-3}$ and temperatures from 150 to
3000$\ \mathrm{K}$, 
the temperatures prevailing in the atmospheres of brown dwarf stars.
These potentials lead to far wing line profiles with a KHe satellite
 on the blue side of the D lines, and a monotonically 
 decreasing wing on the red side (Figs.~\ref{varTD2}-\ref{varTD1}).
  When the line center is strongly saturated,
  these wings can become a significant source of opacity.
  In Figs.~\ref{varTD2} and ~\ref{extension} we highlight the region of
  interest near the KHe quasi-molecular satellite  for comparison with
  previous results described  in \citet{allard2003} and calculations based on the ab initio potentials of \citet{blank2012}.

At this moderate pressure, less than 1~bar, the line profile intensities
are related to the perturbations caused by a single binary collision event. 
 The binary model, for an optically active atom in collision with one perturber,
 is valid for the whole profile except the central part of the line.
The  formation of the  KHe quasi-molecule at 0.707~\,$\mu$m  is
in agreement with quantum mechanical calculations carried out by
\citet{zhu2006} and \citet{alioua2012}.

\begin{figure}
 \centering
\resizebox{0.46\textwidth}{!}
{\includegraphics*{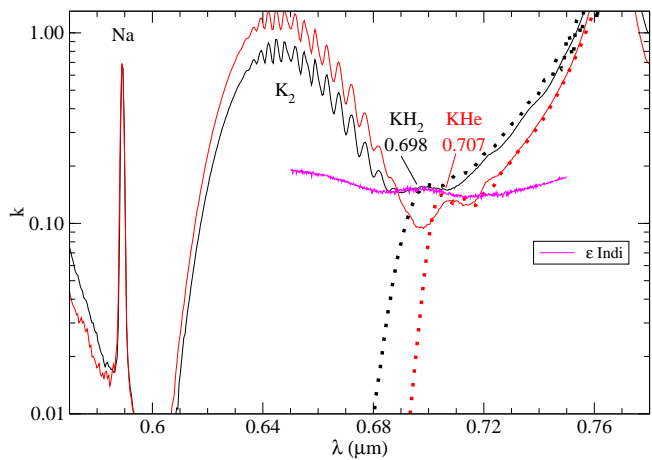}}
\caption  {Comparison of the theoretical blue line wings of
  the resonance line 
  of K (dotted lines) with the experimental spectra (full lines).
  The red curve represents K-He and the black curve represents K-H$_2$.
  The  density of   
  perturbers is 10$^{19}$~cm$^{-3}$ at 800$\ \mathrm{K}$.
   The experimental spectra show the K$_2$ dimer absorption.
  Detail of the spectrum of $\epsilon\,$Indi~Ba shows 
  the blue satellite of the K~resonance lines. (Observational data
  of Mark McCaughrean, as was described in \citet{allard2007b}.)
       }
\label{fig:exp}
\end{figure}

\begin{figure}
  \begin{center} 
    {\resizebox{0.46\textwidth}{!}{
        \includegraphics*[clip]{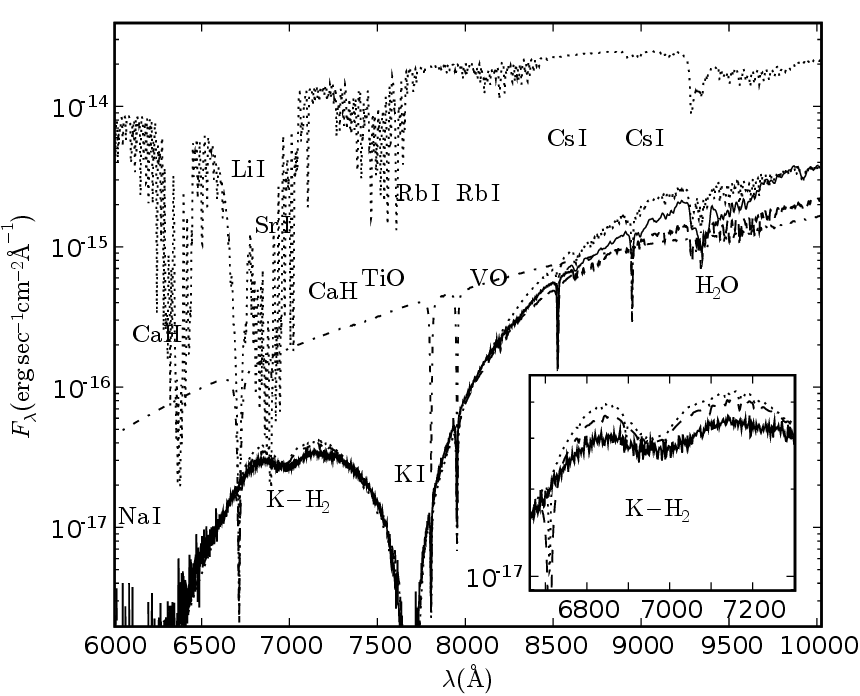} }}
\caption{FORS2 red optical spectrum of the T1 dwarf $\varepsilon$\,Indi\,Ba
(solid), compared to synthetic spectra for a 1.3\,Gyr (dashed) 
and 2\,Gyr (dotted) model (extracted from \citet{allard2007b}).
      \label{fig:eIndi} 
    }
\end{center}
\end{figure}

\begin{figure}
 \centering
\resizebox{0.46\textwidth}{!}
{\includegraphics*{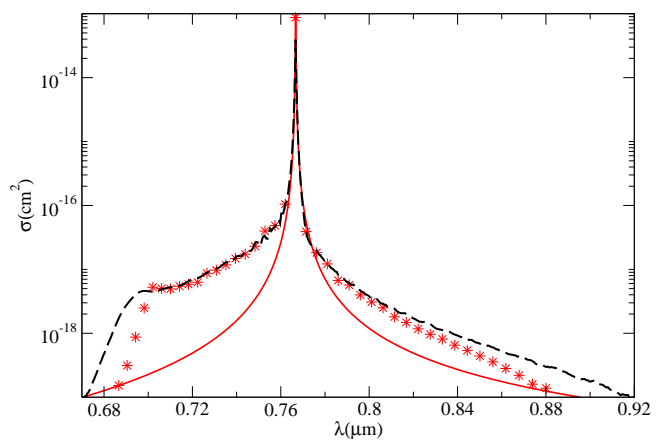}}
\caption  {Absorption cross section 
of  the $D2$  component of the  resonance lines of K perturbed by
He and H$_2$ collisions. The  density   of   
perturbers is   10$^{20}$~cm$^{-3}$ at 1000$\ \mathrm{K}$.
The red curve represents K-He and the black curve represents K-H$_2$.
The Lorentzian approximation is overplotted for comparison.}
\label{fig:KHeH2}
\end{figure}

\subsection{Laboratory spectra of K  perturbed by He and H$_2$}  
\label{sec:exp}

Previous experimental spectra explored the near wing of the alkalies with rare
gases, and supported the development of pseudopotential theories 
for the interactions of these complex systems \citep{mccartan1976,lwin1978}. 
Experiments at high pressures  and room temperatures were carried out 
by~\citet{scheps1975} for Li-He and by~\citet{york1975} for Na-He
but no measurements were done for K-He.
More recently, \citet{shindo2007} reported experimental profiles 
for the second member of the K principle series broadened by He.
 At lower temperatures, experiments to determine the emission spectrum were
conducted by Havey et al. (1980) for Na-He and by Enomoto et al. (2004) for
Li/Na/K-He.
It is a straightforward application of dispersive laboratory spectroscopy to
determine the absorption spectrum of the alkalies broadened by gases, since
modern detector technology enables high-precision quantitative
measurements of the absorption coefficient.
The spectra shown  in Fig.~\ref{fig:exp} were acquired using conventional
absorption spectroscopy of an alkali vapor cell in a uniformly heated oven.
The measurements used a methodology that has been described
previously \citep{kielkopf1980,kielkopf1983}.
In this section, we will briefly describe the experiment design. More
details were given in \citet{allard2012b}.
A CCD detector recorded the full spectral range in a single exposure
with a 0.16~meter focal length Czerny-Turner spectrograph operated at
4~\AA\/ pixel dispersion.
The alkali was introduced as 99.95\% purity metal containing detectable
Na, Rb, and Cs in the case of K, and as K, Rb, and Cs in the case of Na.
Gases were spectroscopic-grade and trace impurities were not significant
for the line broadening measurements.
For Na and K with H$_2$ and other gases, the difficulties lie primarily in
handling the alkalies, and in the usual hazards of H$_2$, but temperatures
up to 1000$\ \mathrm{K}$ and pressures up to one atmosphere are readily obtained.
The laboratory spectrum of K with H$_2$, He, and other gases also shows a 
systematic pattern of satellites in the blue wing 
\citep{kielkopf2017}.  Xe and Kr produce the strongest
satellites closest to the parent line, as was expected from the longer range 
but weaker interactions of these noble gases.  By contrast, H$_2$ and 
He produce satellites farthest from the line. 
The  absorption spectra of K with He and H$_2$ were measured
at pressures under 1~bar under controlled conditions in the laboratory
to compare with the unified theory calculations and validation
of ab initio potentials.  
In previous experimental and theoretical work on Na perturbed by H$_2$, we also
reported the Na line wings measured similarly \citep{allard2012b}.
As in that work, the data we report on here are from a series 
of spectra taken of a K absorption cell
with a uniformly heated central section, 30~cm long by 2.2~cm in diameter.
Both Na and K form stable Na$_2$ and K$_2$ molecules that absorb in part of the
region of interest. The presence of these dimers is unavoidable at the
relatively low temperatures used in an absorption cell. They are apparent
because the vapor pressure of the alkali is intentionally raised to make
the extreme wing of the atomic line observable.
We also took data with Kr buffer gas to remove this molecular
background and reveal only the atomic K-He and K-H$_2$ contributions.  
Spectra with Kr, and all rare gases other than He, have classically inaccessible 
line wing contributions to the
spectral region where the singular KHe and KH$_2$ features occur, and 
therefore provide an  absorption coefficient only
due to K$_2$.  It is straightforward then to remove this component  
by subtraction and leave only the coefficient of absorption due to K-He and
K-H$_2$ in the reduced experimental data.
While the  pseudopotentials of \citet{rossi1985} overestimated the
displacement of the KH$_2$ satellite from the line, ab initio
potentials described in \citet{allard2007a} predicted a satellite that
matches the observations of $\epsilon$Indi Ba more
precisely (Fig.~\ref{fig:eIndi}).
The observed laboratory spectra of K with H$_2$ and He  shown in Fig.~\ref{fig:exp}
 agree exceptionally well with our theoretical profiles.  They confirm the
 identification of the brown dwarf spectral feature of $\epsilon$\,Indi\,Ba,b
 as being due to K--H$_2$/He.

\subsection{Quasi-molecular absorption  in  T-type  brown dwarfs}
\label{sec:Tdwarf}

A broad and shallow absorption feature centered around 0.695\,$\mu$m
in the blue wing of the K doublet at 0.77\,$\mu$m was identified by
\citep{burgasser2003} as the  CaH system. The model atmosphere and
synthetic spectra of  \citet{allard2003} have shown the possible
detectability of the  quasi-molecular line due to K-H$_2$ collisions.
Figure~\ref{fig:eIndi}, extracted from \citet{allard2007b}, shows a 
region of an exceptionally high-quality spectrum of $\epsilon$\,Indi\,Ba,b
that includes the blue wing of the K doublet. The 
\object{$\varepsilon$\,Indi\,Ba,b} binary system is a unique test
for brown dwarf models. This spectrum illustrates how the visible
spectra of brown dwarfs may be dominated by the strong absorption from 
two abundant alkalies, Na and K.

In \citet{allard2007a}, we have shown that ab initio K-H$_2$
potentials systematically less repulsive than the pseudo-potentials 
of \citet{rossi1985} enabled a KH$_2$ quasi-molecular line satellite
to be predicted. This satellite closely matches the position and shape of an
observed feature in the spectrum of the T1 dwarf $\varepsilon$\,{Indi}\,Ba
\citep{allard2007b}.

Figure~10 of \citet{allard2016b} shows an exceptionally good match
between the K-H$_2$ experimental spectrum and the model from the unified
line shape theory with the new  K-H$_2$ potentials. As was noticed in  
\citet{allard2007b}, the new theoretical line satellite apparently does not
extend far enough to the red to reproduce the more elongated
shape of the observed feature (Fig.~\ref{fig:eIndi}). Collisions with
H$_2$ are preponderant in brown dwarf atmospheres with an effective 
temperature of about 1000$\ \mathrm{K}$, but collisions with He should be
considered as well (see Fig.~\ref{fig:KHeH2}), and new K-He collisional
profiles might improve the agreement with the observation of the
spectrum of the T1 dwarf $\varepsilon$\,Indi\,Ba.

In the next section, we examine the dependence of the new KHe line
parameters on the temperature.

\section{Line core parameters}
\label{sec:param}

\begin{figure}
 \centering
\resizebox{0.46\textwidth}{!}
{\includegraphics*{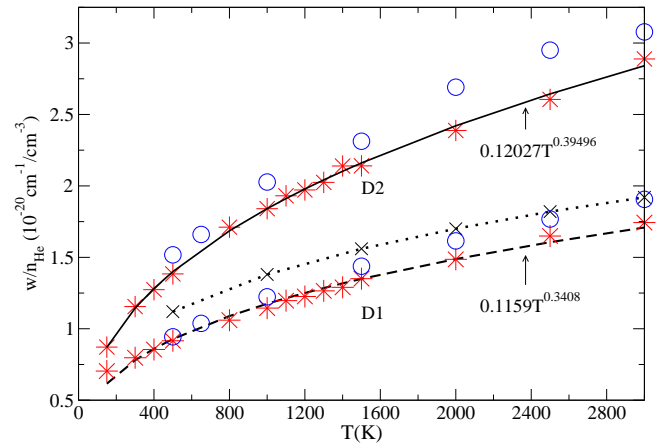}}
\caption  {Variation with temperature of the broadening rate
  ($w$/$n_{\mathrm{He}}$) of the 4$P_{3/2}$-4$s$ (solid line)
  and 4$P_{1/2}$-4$s$ (dashed line) resonance lines of K
  perturbed by He collisions. A comparison is made with \citet{pascale1983}
  (blue circle) and van der Waals (dotted black line). The rates are in
  units of 10$^{-20}$ cm $^{-1}$/cm$^{-3}$.}
\label{fig:wimp}
\end{figure}

\subsection{Semi-classical calculations}

The theory of spectral line shapes, especially the unified approach that we have
developed and refined, makes possible models of stellar spectra 
that account for the centers of spectral lines and their extreme wings in
one consistent treatment.

The impact theories of pressure broadening \citep{baranger1958a,kolb1958}
are based on the assumption of sudden collisions (impacts)  between the
radiator and perturbing atoms, and are valid  when frequency displacements,
\mbox {  $\Delta$ $\omega$ = $\omega$ - $\omega_0$}, and gas densities are
sufficiently  small. In impact broadening, the duration of the collision
is assumed to be small compared to the interval  between collisions, and
the results describe the line within a few line widths of the center.
One outcome of our unified approach is that we may evaluate the difference
between the impact limit and the general unified profile, and establish
with certainty the region of validity of an assumed Lorentzian
profile.  
In the planetary and brown dwarf upper atmospheres, 
the He density is of the order of $10^{16}$ cm$^{-3}$
in the region of line core formation.
At these sufficiently low densities of the perturbers, the symmetric center
of a spectral line is Lorentzian and can be defined by two line 
parameters, the width and the shift of the main line.
These quantities can be obtained in the impact limit
(s $\rightarrow$ $\infty$) of the general
calculation of the autocorrelation function (Eq.~121 of \cite{allard1999}).
In the following discussion, we refer to this line width as measured by half the
full width at half the maximum intensity -- what is customarily termed HWHM.

In \citet{allard2007c}, spectral line widths of the light alkalies 
perturbed by He and H$_2$ were presented for conditions prevailing
in brown dwarf atmospheres. We used pseudo-potentials in a SC
unified theory of the spectral line broadening \citep{allard1999} to
compute line core parameters. For the specific study of the 
$D1$ ($P_{1/2}$) and $D2$ ($P_{3/2}$) components, we need to take 
the spin-orbit coupling of the alkali into account. This is done using
an atom-in-molecule intermediate spin-orbit coupling scheme, analogous
to the one derived by \citet{cohen1974}. The degeneracy is partially
split by the coupling and the distinction between $D1$  and $D2$ results.
We used the molecular structure calculations  performed by~\citet{pascale1983} 
for the adiabatic potentials of alkali-metal--He systems.

The new determinations of line width using our adopted  potentials
in a wide range of temperatures  are presented in Fig.~\ref{fig:wimp}.
The line widths, $w$ (HWHM), are linearly dependent on He density, and
a power law in temperature is given for the $^2P_{1/2}$ component by
\begin{equation}
w = 0.1159 \times 10^{-20} n_{\mathrm{He}} \, T^{0.341}
\label{eq:wfitD1}
\end{equation}
and for the $^2P_{3/2}$ component   by
\begin{equation}
w = 0.1203 \times 10^{-20} n_{\mathrm{He}} \, T^{0.395}
\label{eq:wfitD2}
.\end{equation}

These expressions may be used to compute the widths for temperatures of
stellar atmospheres from 150 to at least 3000~K. When it is assumed that
the main interaction between two atoms is the long-range van der Waals
interaction of two dipoles, the Lindholm-Foley theory gives the usual
formulae for the width and shift. The van der Waals damping constant is
calculated according to the impact theory of the collision broadening.
In Fig.~\ref{fig:wcomp}, we compare our calculation of the 
SC KHe $D1$ and $D2$ broadening rates using two different 
sets of potential energy curves, our adopted potentials, and the 
ab initio potentials from \citet{blank2012}.

\subsection{Quantum calculations}

Quantum scattering calculations of the $D1$ and $D2$ line broadening rates
and shift rates were performed using the Baranger-Lindholm (BL) theory 
\citep{baranger1958a} in the manner described in our previous paper 
for Na-He collisions \citep{allard2023}. Relative phase shifts between
the ground and excited states are obtained from plane wave scattering
calculations, using the relevant potential energy curves for 
fine-structure states involving the two lines. Calculations were
performed over a range of temperatures, with each temperature represented
by an average relative momentum, $\bar{k}$, between radiator and perturber
atoms. Partial wave expansions were used up to maximum angular momenta, 
$l_{\mathrm{max}} = 76 \bar{k}$, where 76~a.u. is the interatomic 
distance over which the K-He potential is significant, for the states 
considered here, and where $\bar{k}$ ranged from 2.3--12.7~a.u. for the 
temperature range, 100--3000~K. Radial solutions of the Schr\"{o}dinger 
equation were computed out to $R_{\mathrm{max}} = 300$~a.u.

\begin{figure}
 \centering
\resizebox{0.46\textwidth}{!}
{\includegraphics*{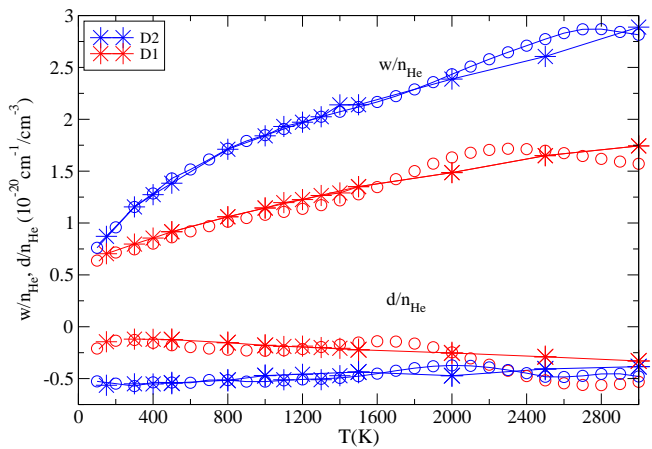}}
\caption  {Variation with temperature of the broadening rate
  ($w$/$n_{\mathrm{He}}$) and shift rate ($d$/$n_{\mathrm{He}}$)
  of the 4$P_{3/2}$-4$s$ (blue) and 4$P_{1/2}$-4$s$ (red)  
  resonance lines of K perturbed by He collisions. The BL theory is represented by circles and the
  SC theory by stars. The rates are in units of 
  10$^{-20}$ cm $^{-1}$/cm$^{-3}$.}
\label{fig:wd}
\end{figure}

Figure~\ref{fig:wd} overlays the quantum calculations of the broadening
and shift rates obtained from BL theory with the rates from the SC
theory. Excellent agreement between the SC and
quantum calculations is seen in the range 150--1500 K. Beyond 1500~K, the
quantum rates exhibit stronger coherent oscillations as a result of 
using the average relative momentum, $\bar{k}$, to represent the temperature. 
Ideally, for a gas in thermal equilibrium, an average of the broadening
and shift rates would be computed using the thermal distribution of
relative momenta,
\begin{equation}
 f(k) = {\left(\frac{\mu}{2\pi k_{\mathrm{B}}T}\right)}^{3/2}
 \exp\left(-\frac{\hbar^2k^2}{2\mu k_{\mathrm{B}}T}\right)
\label{eq:kdist}
,\end{equation}
where $\mu$ is the reduced mass of the radiator-perturber system.
A set of scattering phase shifts, for $l = 0$--$l_{\mathrm{max}}$, must
then be computed at each $k$, over a fine enough mesh in $k$, and
extended over a sufficiently large range in $k$ in order to obtain
an accurate thermal average at each temperature. While we expect that the
oscillations in the BL rates will be smoothed out by performing the
thermal average, we note that the broadening rates for both SC
and quantum calculations, shown in fig.~\ref{fig:wd}, remain close to
each other up to $T = 3000$~K. Thus, we expect that 
equations~\ref{eq:wfitD1}--\ref{eq:wfitD2} will also represent the
broadening rates of the $D1$ and $D2$ lines from BL theory when 
properly averaged over the relative momentum distribution 
(Eqn.~\ref{eq:kdist}) of each temperature.

In Table~\ref{tab:compw} we compare our $D1$ and $D2$
broadening rates from BL theory, computed at 300~K, with other
recently predicted rates, given in Table~4 of \citet{ding2022}.
We also compare the predicted broadening rates with the
recent experimental measurements of \citet{ding2022}. Our
predicted broadening rates are seen to agree to better than
1\% with the experimental rates given in Table~3 of \citet{ding2022},
for both $D1$ and $D2$ lines. Our BL rates at $T = 300$~K
are also in excellent agreement with the quantum calculations of
\citet{mullamphy2007}.

\begin{table}
\caption{Comparison of theoretical and experimental broadening
rates, $w/n_{\mathrm He}$ ($10^{-20}$~cm$^{-1}$/cm$^{-3}$), of
the K resonance lines.}
\label{tab:compw}
\begin{tabular}{lccc}
  \hline
\noalign{\medskip}
\multicolumn{1}{c}{Reference} &
\multicolumn{1}{c}{$D1$} &
\multicolumn{1}{c}{$D2$} &
\multicolumn{1}{l}{T(K)} \\
\hline
\multicolumn{4}{l}{Theory} \\
\hline
\noalign{\medskip}
   {This work (BL)}      & 0.7464 & 1.152  & 300 \\
\noalign{\medskip}
   \citet{ding2022}      & 0.7207 & 0.7244 & 296 \\
\noalign{\medskip}
   \citet{mullamphy2007} & 0.7441 & 1.113  & 296 \\
\noalign{\medskip}
   \citet{blank2014}     & 0.8809 & 1.136  & 296 \\
\hline
\multicolumn{4}{l}{Experiment} \\
\hline
\noalign{\medskip}
   ~\citet{ding2022} & 0.7474 & 1.141  & 296 \\
\hline
\end{tabular}
\end{table}

\subsection{Contribution of the different components}

In Tables~B1 and C1 we report our computed half-width broadening and shift 
rates obtained in the SC theory and in the quantum 
calculation (BL) for the different transitions. The $P_{1/2}$ line is due to
a simple isolated $A$ $\Pi_{1/2}$ state, whereas the $P_{3/2}$ line comes
from the $A$ $\Pi_{3/2}$ and $B$ $\Sigma_{1/2}$ adiabatic states arising from
the $3p$ $P_{3/2}$ atomic state. The broadening of the $B$ $\Sigma_{1/2}$  
state is most sensitive to potential at intermediate separations, where 
the potential starts to deviate from the long range part. This result 
confirms the study by \citet{roueff1969} and \citet{lortet1969}  of the
collisions with light atoms whose polarisability is small. It was shown
by them that the width of spectral lines due to collisions with hydrogen
atoms does not arise from the van der Waals dispersion forces but from a
shorter-range interaction.

\section{Conclusions}
The optical spectra of L- and T-dwarfs exhibit a continuum 
dominated by the far wings of the absorption profiles 
of the Na $3s-3p$ and K $4s-4p$ doublets perturbed by molecular 
hydrogen and helium.
Studies of observed L- and T-dwarf spectra by \citet{liebert2000} and
\citet{burrows2001}  showed clearly the importance of extended
K line wings and pointed out the need for
spectral broadening calculations more accurate than Lorentzian profiles.
Understanding the shape of these lines is essential 
to modeling the transport of radiation from the interior.
Compared to the commonly used van der Waals broadening in the
impact approximation, major first improvements in the theoretical description of pressure broadening
have been made by \citet{burrows2003} and \citet{allard2003}.

All line shape characteristics depend on the accuracy of the  atomic
potentials and transition moments, and laboratory observations are crucial
for testing the molecular data.
Spectral line shapes shown here were computed in a unified theory \citep{allard1999}
using  ab initio potentials that reproduce laboratory
and astrophysical data
\citep{allard2012b,allard2016b,allard2019,allard2023}.
We have shown that they  confirm the 
identification of a brown dwarf spectral feature is due to line satellites of potassium perturbed by
H$_2$. This feature was previously interpreted
as CaH absorption bands. An additional contribution from the new  K--He opacities reported here should improve
the agreement.
These dwarf star spectroscopic measurements required the largest available ground-based
telescopes with complex and efficient spectrographs to acquire data
with an adequate signal-to-noise ratio.
 Consequently, examples of  high-quality spectra of brown dwarfs for wavelengths below  the K resonance lines are rare.
  Indeed, this spectral region from 0.68 to 0.74~$\mu$m  is very
  temperature-dependent and is very constraining on atmosphere models
  in comparison to observations 
because of the overlap of the red wing of
the (3$s$-3$p$) resonance line of sodium  with the blue wing of the
(4$s$-4$p$) resonance line of potassium \citep{allard2007a}.
The contribution of the quasi-molecular KHe absorption near 0.707~$\mu$m
that we find is expected to reproduce
the observations of the T-dwarf $\varepsilon$\,Indi\,Ba and similar stars in a more realistic 
way than previous calculations did \citep{allard2007b}.
The  new opacity tables of K--He will be archived at the CDS.

\bibliographystyle{aa} 

\begin{thebibliography}{57}
\expandafter\ifx\csname natexlab\endcsname\relax\def\natexlab#1{#1}\fi

\bibitem[{Allard {et~al.}(2007{\natexlab{a}})Allard, Allard, Homeier, Kielkopf,
  McCaughrean, \& Spiegelman}]{allard2007b}
Allard, F., Allard, N.~F., Homeier, D., {et~al.} 2007{\natexlab{a}}, A\&A, 474,
  L21

\bibitem[{Allard {et~al.}(2001)Allard, Hauschildt, Alexander, Tamanai, \&
  A.Schweitzer}]{allard2001}
Allard, F., Hauschildt, P.~H., Alexander, D.~R., Tamanai, A., \& A.Schweitzer.
  2001, Ap J., 556, 357

\bibitem[{Allard {et~al.}(2003)Allard, Allard, Hauschildt, Kielkopf, \&
  Machin}]{allard2003}
Allard, N.~F., Allard, F., Hauschildt, P.~H., Kielkopf, J.~F., \& Machin, L.
  2003, A\&A, 411, L473

\bibitem[{Allard {et~al.}(2005)Allard, Allard, \& Kielkopf}]{allard2005}
Allard, N.~F., Allard, F., \& Kielkopf, J.~F. 2005, A\&A, 440, 1195

\bibitem[{Allard \& Kielkopf(1982)}]{allard1982}
Allard, N.~F. \& Kielkopf, J.~F. 1982, Rev. Mod. Phys., 54, 1103

\bibitem[{Allard {et~al.}(2007{\natexlab{b}})Allard, Kielkopf, \&
  Allard}]{allard2007c}
Allard, N.~F., Kielkopf, J.~F., \& Allard, F. 2007{\natexlab{b}}, EPJ D, 44,
  507

\bibitem[{{Allard} {et~al.}(2023){Allard}, {Myneni}, {Blakely}, \&
  {Guillon}}]{allard2023}
{Allard}, N.~F., {Myneni}, K., {Blakely}, J.~N., \& {Guillon}, G. 2023, \aap,
  674, A171

\bibitem[{Allard {et~al.}(1999)Allard, Royer, Kielkopf, \&
  Feautrier}]{allard1999}
Allard, N.~F., Royer, A., Kielkopf, J.~F., \& Feautrier, N. 1999, Phys. Rev. A,
  60, 1021

\bibitem[{Allard \& Spiegelman(2006)}]{allard2006}
Allard, N.~F. \& Spiegelman, F. 2006, A\&A, 452, 351

\bibitem[{Allard {et~al.}(2007{\natexlab{c}})Allard, Spiegelman, \&
  Kielkopf}]{allard2007a}
Allard, N.~F., Spiegelman, F., \& Kielkopf, J.~F. 2007{\natexlab{c}}, A\&A,
  465, 1085

\bibitem[{{Allard} {et~al.}(2016){Allard}, {Spiegelman}, \&
  {Kielkopf}}]{allard2016b}
{Allard}, N.~F., {Spiegelman}, F., \& {Kielkopf}, J.~F. 2016, A\&A, 589, A21

\bibitem[{Allard {et~al.}(2012)Allard, Spiegelman, Kielkopf, Tinetti, \&
  Beaulieu}]{allard2012b}
Allard, N.~F., Spiegelman, F., Kielkopf, J.~F., Tinetti, G., \& Beaulieu, J.~P.
  2012, A\&A, 543, A159

\bibitem[{{Allard} {et~al.}(2019){Allard}, {Spiegelman}, {Leininger}, \&
  {Molliere}}]{allard2019}
{Allard}, N.~F., {Spiegelman}, F., {Leininger}, T., \& {Molliere}, P. 2019,
  \aap, 628, A120

\bibitem[{Baranger(1958)}]{baranger1958a}
Baranger, M. 1958, Phys. Rev., 111, 481

\bibitem[{{Blank} \& {Weeks}(2014)}]{blank2014}
{Blank}, L. \& {Weeks}, D.~E. 2014, \pra, 90, 022510

\bibitem[{{Blank} {et~al.}(2012){Blank}, {Weeks}, \& {Kedziora}}]{blank2012}
{Blank}, L., {Weeks}, D.~E., \& {Kedziora}, G.~S. 2012, \jcp, 136, 124315

\bibitem[{{Blouin} {et~al.}(2019){Blouin}, {Dufour}, {Allard}, {Salim}, {Rich},
  \& {Koopmans}}]{blouin2019c}
{Blouin}, S., {Dufour}, P., {Allard}, N.~F., {et~al.} 2019, \apj, 872, 188

\bibitem[{{Boutarfa} {et~al.}(2012){Boutarfa}, {Alioua}, {Bouledroua},
  {Allouche}, \& {Aubert-Fr{\'e}con}}]{alioua2012}
{Boutarfa}, H., {Alioua}, K., {Bouledroua}, M., {Allouche}, A.~R., \&
  {Aubert-Fr{\'e}con}, M. 2012, \pra, 86, 052504

\bibitem[{Burgasser {et~al.}(2003)Burgasser, Kirkpatrick, Liebert, \&
  Burrows}]{burgasser2003}
Burgasser, A.~J., Kirkpatrick, J.~D., Liebert, J., \& Burrows, A. 2003, ApJ,
  594, 510

\bibitem[{{Burningham} {et~al.}(2017){Burningham}, {Marley}, {Line}, {Lupu},
  {Visscher}, {Morley}, {Saumon}, \& {Freedman}}]{marley2017}
{Burningham}, B., {Marley}, M.~S., {Line}, M.~R., {et~al.} 2017, \mnras, 470,
  1177

\bibitem[{{Burrows} {et~al.}(2002){Burrows}, {Burgasser}, {Kirkpatrick},
  {Liebert}, {Milsom}, {Sudarsky}, \& {Hubeny}}]{burrows2002}
{Burrows}, A., {Burgasser}, A.~J., {Kirkpatrick}, J.~D., {et~al.} 2002, \apj,
  573, 394

\bibitem[{Burrows {et~al.}(2001)Burrows, Hubbard, Lunine, \&
  Liebert}]{burrows2001}
Burrows, A., Hubbard, W.~B., Lunine, J.~I., \& Liebert, J. 2001, Rev. Mod.
  Phys., 73, 719

\bibitem[{Burrows {et~al.}(2000)Burrows, M.~S.~Marley, \& Sharp}]{burrows2000}
Burrows, A., M.~S.~Marley, M.~S., \& Sharp, C.~M. 2000, ApJ, 531, 438

\bibitem[{Burrows \& Volobuyev(2003)}]{burrows2003}
Burrows, A. \& Volobuyev, M. 2003, ApJ, 583, 985

\bibitem[{{Chubb} {et~al.}(2021){Chubb}, {Rocchetto}, {Yurchenko}, {Min},
  {Waldmann}, {Barstow}, {Molli{\`e}re}, {Al-Refaie}, {Phillips}, \&
  {Tennyson}}]{chubb2021}
{Chubb}, K.~L., {Rocchetto}, M., {Yurchenko}, S.~N., {et~al.} 2021, \aap, 646,
  A21

\bibitem[{Cohen \& Schneider(1974)}]{cohen1974}
Cohen, J.~S. \& Schneider, B. 1974, J. Chem. Phys., 61, 3230

\bibitem[{{Ding} {et~al.}(2022){Ding}, {Vandervort}, {Freedman}, {Strand},
  {Marley}, \& {Hanson}}]{ding2022}
{Ding}, Y., {Vandervort}, J.~A., {Freedman}, R.~S., {et~al.} 2022, \jqsrt, 283,
  108149

\bibitem[{Enomoto {et~al.}(2004)Enomoto, Hirano, Kumakura, Takahashi, \&
  Yabuzaki}]{enomoto2004}
Enomoto, K., Hirano, K., Kumakura, M., Takahashi, Y., \& Yabuzaki, T. 2004,
  Phys. Rev. A, 69, 012501

\bibitem[{Gonzales {et~al.}(2021)Gonzales, Burningham, Faherty, Visscher,
  Marley, Lupu, Freedman, \& Lewis}]{freedman2021}
Gonzales, E.~C., Burningham, B., Faherty, J.~K., {et~al.} 2021, The
  Astrophysical Journal, 923, 19

\bibitem[{Havey {et~al.}(1980)Havey, Frolking, \& Wright}]{havey1980}
Havey, M.~D., Frolking, S.~E., \& Wright, J.~J. 1980, Phys. Rev. Lett., 45,
  1783

\bibitem[{{Hou Yip} {et~al.}(2020){Hou Yip}, {Changeat}, {Edwards}, {Morvan},
  {Chubb}, {Tsiaras}, {Waldmann}, \& {Tinetti}}]{changeat2020}
{Hou Yip}, K., {Changeat}, Q., {Edwards}, B., {et~al.} 2020, in European
  Planetary Science Congress, EPSC2020--67

\bibitem[{Kielkopf(1980)}]{kielkopf1980}
Kielkopf, J.~F. 1980, J. Phys. B: At. Mol. Opt. Phys., 13, 3813

\bibitem[{Kielkopf(1983)}]{kielkopf1983}
Kielkopf, J.~F. 1983, J. Phys. B: At. Mol. Opt. Phys., 16, 3149

\bibitem[{Kielkopf {et~al.}(2017)Kielkopf, Allard, Alekseev, Spiegelman,
  Guillon, \& Berriche}]{kielkopf2017}
Kielkopf, J.~F., Allard, N.~F., Alekseev, V.~A., {et~al.} 2017, Journal of
  Physics: Conference Series, 810, 012023

\bibitem[{{Kielkopf} {et~al.}(2012){Kielkopf}, {Allard}, \&
  {Babb}}]{kielkopf2012}
{Kielkopf}, J.~F., {Allard}, N.~F., \& {Babb}, J. 2012, in EAS Publications
  Series, Vol.~58, EAS Publications Series, 75--78

\bibitem[{{Kolb} \& {Griem}(1958)}]{kolb1958}
{Kolb}, A.~C. \& {Griem}, H. 1958, Physical Review, 111, 514

\bibitem[{{Lacy} \& {Burrows}(2023)}]{lacy2023}
{Lacy}, B. \& {Burrows}, A. 2023, \apj, 950, 8

\bibitem[{Liebert {et~al.}(2000)Liebert, Reid, Burrows, Burgasser, Kirpatrick,
  \& Gizis}]{liebert2000}
Liebert, J., Reid, I., Burrows, A., {et~al.} 2000, ApJ, 533, L155

\bibitem[{{Lortet} \& {Roueff}(1969)}]{lortet1969}
{Lortet}, M.~C. \& {Roueff}, E. 1969, \aap, 3, 462

\bibitem[{Lwin \& McCartan(1978)}]{lwin1978}
Lwin, N. \& McCartan, D.~G. 1978, J. Phys. B: At. Mol. Opt. Phys., 11, 3841

\bibitem[{McCartan \& Farr(1976)}]{mccartan1976}
McCartan, D.~G. \& Farr, J.~M. 1976, J. Phys. B: At. Mol. Opt. Phys., 9, 985

\bibitem[{Mullamphy {et~al.}(2007)Mullamphy, Peach, Venturi, Whittingham, \&
  Gibson}]{mullamphy2007}
Mullamphy, D. F.~T., Peach, G., Venturi, V., Whittingham, I.~B., \& Gibson,
  S.~J. 2007, J. Phys. B: At. Mol. Opt. Phys., 40, 1141

\bibitem[{Nakayama \& Yamashita(2001)}]{nakayama2001b}
Nakayama, A. \& Yamashita, K. 2001, J. Chem. Phys., 114, 780

\bibitem[{{Nikolov} {et~al.}(2022){Nikolov}, {Sing}, {Spake}, {Smalley},
  {Goyal}, {Mikal-Evans}, {Wakeford}, {Rustamkulov}, {Deming}, {Fortney},
  {Carter}, {Gibson}, \& {Mayne}}]{nikolov2022}
{Nikolov}, N.~K., {Sing}, D.~K., {Spake}, J.~J., {et~al.} 2022, \mnras, 515,
  3037

\bibitem[{{Oreshenko} {et~al.}(2020){Oreshenko}, {Kitzmann},
  {M{\'a}rquez-Neila}, {Malik}, {Bowler}, {Burgasser}, {Sznitman}, {Fisher}, \&
  {Heng}}]{oreshenko2020}
{Oreshenko}, M., {Kitzmann}, D., {M{\'a}rquez-Neila}, P., {et~al.} 2020, \aj,
  159, 6

\bibitem[{Pascale(1983)}]{pascale1983}
Pascale, J. 1983, Phys. Rev. A, 28, 632

\bibitem[{{Phillips} {et~al.}(2020){Phillips}, {Tremblin}, {Baraffe},
  {Chabrier}, {Allard}, {Spiegelman}, {Goyal}, {Drummond}, \&
  {H{\'e}brard}}]{phillips2020}
{Phillips}, M.~W., {Tremblin}, P., {Baraffe}, I., {et~al.} 2020, \aap, 637, A38

\bibitem[{Rossi \& Pascale(1985)}]{rossi1985}
Rossi, F. \& Pascale, J. 1985, Phys. Rev. A, 32, 2657

\bibitem[{{Roueff} \& {van Regemorter}(1969)}]{roueff1969}
{Roueff}, E. \& {van Regemorter}, H. 1969, \aap, 1, 69

\bibitem[{{Samra} {et~al.}(2023){Samra}, {Helling}, {Chubb}, {Min}, {Carone},
  \& {Schneider}}]{chubb2023}
{Samra}, D., {Helling}, C., {Chubb}, K.~L., {et~al.} 2023, \aap, 669, A142

\bibitem[{Santra \& Kirby(2005)}]{santra2005}
Santra, R. \& Kirby, K. 2005, J. Chem. Phys., 123, 214309

\bibitem[{Scheps {et~al.}(1975)Scheps, Ottinger, York, \&
  Gallagher}]{scheps1975}
Scheps, R., Ottinger, C., York, G., \& Gallagher, A. 1975, J. Chem. Phys., 63,
  2581

\bibitem[{Shindo {et~al.}(2007)Shindo, Babb, Kirby, \& Yoshino}]{shindo2007}
Shindo, F., Babb, J.~F., Kirby, K., \& Yoshino, K. 2007, J. Phys. B: At. Mol.
  Opt. Phys., 40, 2841

\bibitem[{Szudy \& Baylis(1975)}]{szudy1975}
Szudy, J. \& Baylis, W. 1975, J. Quant. Spectrosc. Radiat. Transfer, 15, 641

\bibitem[{Szudy \& Baylis(1996)}]{szudy1996}
Szudy, J. \& Baylis, W. 1996, Physics Reports, 266, 127

\bibitem[{York {et~al.}(1975)York, Scheps, \& Gallagher}]{york1975}
York, G., Scheps, R., \& Gallagher, A. 1975, J. Chem. Phys., 63, 1052

\bibitem[{Zhu {et~al.}(2006)Zhu, Babb, \& Dalgarno}]{zhu2006}
Zhu, C., Babb, J.~F., \& Dalgarno, A. 2006, Phys. Rev. A, 73, 012506

\end{thebibliography}

\begin{appendix}

  \onecolumn

\section{Extension of the wings}
  
\begin{figure}[h!]
 \centering
\resizebox{0.46\textwidth}{!}
{\includegraphics*{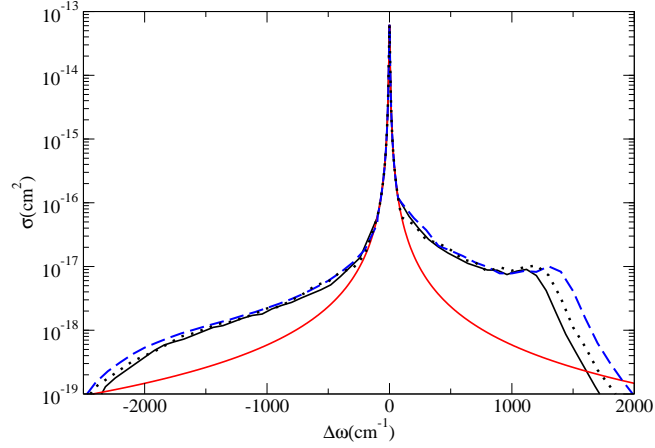}}
\caption{Absorption cross section of the K $D2$ component at 
 $T$  =3000~K, $n_{\mathrm{He}}$=1$\times$~$10^{20}$~cm$^{-3}$ (black line) 
  compared to the Lorentzian profile (red line).
The corresponding  profiles for T=3000~$\ \mathrm{K}$ using
the pseudo-potentials of \citet{pascale1983} (dashed blue line)
and the ab initio potentials of \citet{blank2012} (dotted black line)
are  overplotted.}
\label{extension} 
\end{figure}
  
For a He density lower than n$_{\mathrm{He}}$= 10$^{20}$ atoms~cm$^{-3}$, the core
of the line is described adequately by a Lorentzian profile. This figure  demonstrates that unified  profiles provide more absorption close to the line core, while less far out in the red
wing, compared to the use of a Lorentz profile.

 \begin{center}  
     
\section{Line broadening rate}
  
{Table B1 Computed broadening rates, $w/n_{\mathrm{He}}$ 
 (10$^{-20}$ cm$^{-1}$/cm$^{-3}$),
  of K resonance lines perturbed by He collisions. Values are given from
  both SC and quantum (BL) calculations. }
\label{tab:wKHe}
\begin{flushleft}
\begin{tabular}{lllllllllll}
\noalign{\bigskip}
\hline
\noalign{\bigskip}
\multicolumn{1}{l}{Component} &
\multicolumn{1}{l}{Transition} &
\multicolumn{1}{l}{$w/n_{\mathrm{He}}$} &
\multicolumn{1}{l}{500~K} &
\multicolumn{1}{l}{800~K} &
\multicolumn{1}{l}{1000~K} &
\multicolumn{1}{l}{1500~K} & 
\multicolumn{1}{l}{2000~K} & 
\multicolumn{1}{l}{2500~K} & 
\multicolumn{1}{l}{3000~K} &
\multicolumn{1}{l}{weight}\\
\noalign{\bigskip}
\hline
\noalign{\bigskip}
 $4s$ $^2S_{1/2}$-$4p$ $^2P_{1/2}$ & $A$ $\Pi_{1/2}$-$X$ & sc
& 0.916 & 1.06 & 1.145 & 1.35 & 1.49 & 1.65 & 1.74 & 1\\
\noalign{\medskip}
\noalign{\medskip}
  && BL
& 0.863 & 1.01 & 1.079 & 1.28 & 1.63 & 1.70 & 1.57 & 1\\
\noalign{\medskip}
\noalign{\medskip}
  $4s$ $^2S_{1/2}$-$4p$ $^2P_{3/2}$  &  $A$ $\Pi_{3/2}$-$X$ & sc
& 0.514 & 0.618 & 0.677 & 0.788 & 0.872 & 0.953 & 1.017 & 0.5\\
\noalign{\medskip}
\noalign{\medskip}
  && BL
& 0.535 & 0.638 & 0.684 & 0.731 & 0.894 & 1.055 & 1.034 & 0.5\\
\noalign{\medskip}
\noalign{\medskip}
   & $B$ $\Sigma_{1/2}$-$X$ & sc
& 0.869 & 1.093 & 1.164 & 1.352 & 1.516 & 1.652 & 1.872 & 0.5\\
\noalign{\medskip}
\noalign{\medskip}
&  & BL 
& 0.895 & 1.078 & 1.164 & 1.387 & 1.540 & 1.718 & 1.780 & 0.5\\
\noalign{\medskip}
\noalign{\medskip}
& $w/n_{\mathrm{He}}$ $^2P_{3/2}$ & sc
& 1.384 & 1.711 & 1.841 & 2.140 & 2.388 & 2.605 & 2.889 & \\
\noalign{\medskip}
\noalign{\medskip}
&& BL
& 1.430 & 1.716 & 1.847 & 2.118 & 2.434 & 2.773 & 2.814 & \\
\noalign{\medskip}
\noalign{\medskip}
\hline
\end{tabular}
\\
\end{flushleft}

\newpage

\section{Line shift rate}

{Table C1 Computed shift rates, $d/n_{\mathrm{He}}$ 
  (10$^{-20}$ cm$^{-1}$/cm$^{-3}$),
  of K resonance lines perturbed by He collisions. Values are given
  from both SC and quantum (BL) calculations. }
\label{tab:dKHe}
\begin{flushleft}
\begin{tabular}{lllllllllll}
\noalign{\bigskip}
\hline
\noalign{\bigskip}
\multicolumn{1}{l}{Component} &
\multicolumn{1}{l}{Transition} &
\multicolumn{1}{l}{$d/n_{\mathrm{He}}$} &
\multicolumn{1}{l}{500~K} &
\multicolumn{1}{l}{800~K} &
\multicolumn{1}{l}{1000~K} &
\multicolumn{1}{l}{1500~K} & 
\multicolumn{1}{l}{2000~K} & 
\multicolumn{1}{l}{2500~K} & 
\multicolumn{1}{l}{3000~K} &
\multicolumn{1}{l}{weight}\\
\noalign{\bigskip}
\hline
\noalign{\bigskip}
 $4s$ $^2S_{1/2}$-$4p$ $^2P_{1/2}$ &  $A$ $\Pi_{1/2}$-$X$ & sc
& -0.1245 & -0.154 & -0.180 & -0.222 & -0.253 & -0.290 & -0.330 & 1\\
\noalign{\medskip}
\noalign{\medskip}
  && BL
& -0.1784 & -0.221 & -0.231 & -0.152 & -0.247 & -0.517 & -0.531 & 1\\
\noalign{\medskip}
\noalign{\medskip}
  $4s$ $^2S_{1/2}$-$4p$ $^2P_{3/2}$  & $A$ $\Pi_{3/2}$-$X$ & sc
& -0.496 & -0.534 & -0.555 & -0.592 & -0.640 & -0.660 & -0.703 & 0.5\\
\noalign{\medskip}
\noalign{\medskip}
& & BL
& -0.514 & -0.560 & -0.594 & -0.613 & -0.568 & -0.700 & -0.828 & 0.5\\
\noalign{\medskip}
\noalign{\medskip}
   & $B$ $\Sigma_{1/2}$-$X$ & sc
& -0.047 & 0.018 & 0.084 & 0.156 & 0.168 & 0.253 & 0.318 & 0.5\\
\noalign{\medskip}
\noalign{\medskip}
&  & BL
& -0.023 & 0.045 & 0.065 & 0.136 & 0.194 & 0.226 & 0.349 & 0.5\\
\noalign{\medskip}
\noalign{\medskip}
& $d/n_{\mathrm{He}}$ $^2P_{3/2}$ & sc
& -0.543 & -0.516 & -0.471 & -0.436 & -0.472 & -0.407 & -0.385\\
\noalign{\medskip}
\noalign{\medskip}
&& BL
& -0.538 & -0.515 & -0.529 & -0.477 & -0.373 & -0.482 & -0.479\\
\noalign{\medskip}
\noalign{\medskip}
\hline
\end{tabular}
\\
\end{flushleft}

\end{center}

\section{Comparison of line broadening for two sets of ab initio potentials}

\begin{figure}[h!]
 \centering
\resizebox{0.46\textwidth}{!}
{\includegraphics*{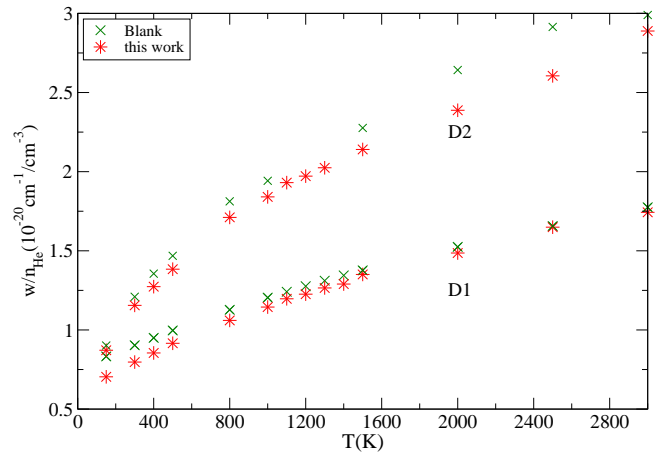}}
\caption  {Variation with temperature of the
  broadening rate  ($w$/$n_{\mathrm{He}}$)
  of the 4$P_{3/2}$-4$s$ ($D$2) and 4$P_{1/2}$-4$s$ ($D$1)  
resonance lines of K perturbed by He collisions. 
	\citet{blank2012} is shown in green and this work (SC) is shown in red.  
 The rates are in units of 10$^{-20}$ cm $^{-1}$/cm$^{-3}$.}
\label{fig:wcomp}
\end{figure}

\end{appendix}

\end{document}